# A New Insight into Land Use Classification Based on Aggregated Mobile Phone Data

TAO PEI
*State Key Laboratory of Resources and Environmental Information System, Institute of Geographic Sciences and Natural Resources Research, CAS, 11A, Datun Road Anwai, Beijing 100101, China*

*SENSEable City Lab, School of Architecture and Planning, Massachusetts Institute of Technology, Cambridge, MA 02139, USA*

Stanislav Sobolevsky (Corresponding Author)
*Senseable City Lab, School Of Architecture And Planning, Massachusetts Institute Of Technology, Cambridge, Ma 02139, Usa*

Carlo Ratti

*Senseable City Lab, School Of Architecture And Planning, Massachusetts Institute Of Technology, Cambridge, Ma 02139, Usa*

Shih-Lung Shaw
*Department Of Geography, University Of Tennessee, 304 Burchfiel Geography Building, Knoxville, Tn 37996-0925, Usa*
*State Key Laboratory Of Information Engineering In Surveying, Mapping And Remote Sensing, Wuhan University, Wuhan, Hubei 430079, China*

Chenghu Zhou
*State Key Laboratory of Resources and Environmental Information System, Institute of Geographic Sciences and Natural Resources Research, CAS, 11A, Datun Road, Beijing 100101, China*

Abstract

Land use classification is essential for urban planning. Urban land use types can be differentiated either by their physical characteristics (such as reflectivity and texture) or social functions. Remote sensing techniques have been recognized as a vital method for urban land use classification because of their ability to capture the physical characteristics of land use. Although significant progress has been achieved in remote sensing methods designed for urban land use classification, most techniques focus on physical characteristics, whereas knowledge of social functions is not adequately used. Owing to the wide usage of mobile phones, the activities of residents, which can be retrieved from the mobile phone data, can be determined in order to indicate the social function of land use. This could bring about the opportunity to derive land use information from mobile phone data. To verify the application of this new data source to urban land use classification, we first construct a time series of aggregated mobile phone data to characterize land use types. This time series is composed of two aspects: the hourly relative pattern, and the total call volume. A semi-supervised fuzzy c-means clustering approach is then applied to infer the land use types. The method is validated using mobile phone data collected in Singapore. Land use is determined with a detection rate of 58.03%. An analysis of the land use classification results shows that the accuracy decreases as the heterogeneity of land use increases, and increases as the density of cell phone towers increases.

Keywords: land use; mobile phone data; classification; FCM; Singapore

## 1. Introduction

The classification of urban land use is essential for urban planning. Urban land use, defined as the recognized human use of land in a city, can be differentiated either by its physical characteristics (such as reflectivity and texture) or social functions (i.e., residential areas are for living whereas industrial areas are for working). Among urban land use classification methods, remote sensing techniques are recognized as a vital method because of their ability to capture the physical characteristics of land use. Conventional land-use remote sensing methods classify land use based on spectral and textual characteristics (Gong and Howarth 1990; Fisher 1997; Shaban and Dikshit 2001; Lu and Weng 2006). Nevertheless, because land use classes are heterogeneous in both their spectral and textural characteristics, methods that rely on remote sensing information and their derived characteristics are unable to differentiate between some land use types (i.e., residential and commercial). Because of this, more auxiliary information, such as contextual properties, field sizes and shapes, parcel information, and expert knowledge, has been used to infer land use patterns (De Wit and Clevers, 2004; Platt and Rapoza, 2008; Wu et al. 2009; Hu and Wang, 2013). However, this need for additional information not only increases the cost, but also delays the update process. Although significant progress has been made in remote sensing techniques, there is a tendency to focus on the utilization of information concerning physical characteristics of land use, and knowledge of social functions is not adequately used in the classification process.

Owing to the wide usage of mobile phones, the daily activities of residents in various regions can be easily captured and used to indicate the social function of the land use type. In other words, within different land use areas, people may demonstrate different routine activities (for example, in residential areas, people usually leave home for work in the morning and return in the evening, whereas in business areas the opposite pattern can be found). This may allow us to derive the activities of residents, and then the social functions of different land use types, from mobile phone data. As a result, mobile phone data may provide a new insight into traditional urban land use from the perspective of social function. The objective of this paper is to verify the applicability of the potential data source for urban land use classification, and then evaluate the results given by this new source of information.

The remainder of the paper is structured as follows. Section 2 introduces a newly constructed time series, as well as the semi-supervised cluster method for urban land use classification. In Section 3, the mobile phone data used in this paper are described. Section 4 presents the overall procedure and the results of land use classification. Section 5 validates the classification result by comparing it with that given by either the call pattern or call volume alone. Section 6 discusses the factors affecting the uncertainty in the classification, and Section 7 presents our conclusions and suggestions for future work relating to land use classification based on mobile phone data.

## 2. Related work

The retrieval of land use from mobile phone data can be divided into two stages. The first is to retrieve the residents' activities based on mobile phone data. The second is to infer land use from the residents' activities. Regarding the first stage, recent research can be grouped into two categories. The first aims to reveal individual mobility patterns using call detail record data, which consist of the different base transceiver station (BTS) locations from which users have made calls (Gonzalez, et al., 2008; Song et al. 2010; Calabrese et al., 2011). The second is based on the aggregation of the total calling time (or numbers) at each BTS in a certain temporal interval. Since our paper only uses the relationship between the mobility and the aggregated mobile phone data in the inference of urban land use, the literature review below will focus on the achievements of aggregated mobile phone data.

The spatiotemporal variation regarding BTS has been extensively studied to retrieve various residents' activities. Recent approaches include the description of urban landscapes (i.e., the space-time structure of residents' activities in a city) (Ratti et al. 2006; Pulselli et al., 2006; Sevtsuk and Ratti, 2010; Sun et al. 2011; Jacobs-Crisioni and Koomen, 2012; Loibl and Peters-Anders, 2012), population estimates (Vieira et al. 2010; Manfredini et al., 2011; Rubioa et al., 2013), the identification of specific social groups (Vaccari et al. 2009), and the detection of social events (Traag et al. 2011; Laura et al. 2012).

The inference of land use types in this context is dependent on their social functions which can be derived from the residents' activities (namely, the overall characteristics of human communication in the urban area). This contains two aspects: the relative weekly calling pattern ("pattern" hereafter) and the total calling volume ("volume" hereafter). The pattern is defined as the share of hourly calling volume in a certain period. The calling volume of a BTS is defined as the total time (or number) of calls managed by that BTS in its area of coverage over a given period of time. Unlike the static residential population density, the volume is the overall characteristic of how many people actually use mobile phones, indicating the activeness of their communicational interactions. To identify and extract recurring patterns of mobile phone usage and relate them to some land use types, Reades et al. (2009) proposed the eigen-decomposition method, a process similar to factoring but suitable for complex datasets. Calabrese et al. (2010) used an eigen-decomposition analysis to reveal the relationship between mobile phone data and the residential and business areas. Caceres et al. (2012) used a new tessellation technique to differentiate parks from residential areas by detecting changes in human density retrieved from mobile phone data.

Although these studies have addressed the relationship between land use and mobile phone data, they have only focused on the identification of specific land use types, not the classification of urban land use. In order to enhance the land use classification, Soto and Frias-Martinez (2011a and 2011b) used the normalized time series of the volume for a weekday and a weekend day (a time series consists of 48 points, each of which is the volume calculated at each hour and normalized by the total volume of the 2 days) to identify the land use pattern. The same method was applied to Twitter data by Frias-Martinez et al. (2012). Andrienko et al. (2013) used the normalized timelines of mobile phone calls at each BTS to identify the heterogeneity of the Ivory Coast at the country scale. Because the normalized data only cover the temporal variation of the volume within the same BTS, the difference in the total volume between BTSs was neglected. Therefore, regarding the problem of heterogeneous land use (for example, downtown areas may have a variety of commercial, residential, and recreational activities), methods based solely on normalized patterns might fail to discern between different land use types that are not homogenous.

To adapt the mobile phone data to urban land use classification, Toole et al. (2012) proposed a supervised classification method for the data that combined the normalized calling pattern and the volume (namely, “activity” in their paper). The aggregated data were first converted to the residual of the Z-score normalization, which reveals the flow into and out of the city center over the course of a day. The random forest method, proposed by Breiman (2001), was then employed to determine land use types. Although this method significantly enhanced the land use classification, two aspects still need to be improved. First, the random forest, similar to the neural network method, is a black box model (Berthold, 2010), which makes the classification difficult to interpret. Second, only two-day pattern (an average weekday and an average weekend) was used to infer the urban land use (Toole et al., 2012). The difference between weekdays and that between weekends are neglected, despite the fact that the significant differences exist between weekdays and between weekends in terms of activities of residents (Jia and Jiang, 2012; Liu et al., 2012; Soto and Frias-Martinez, 2011a).

Although previous studies have made substantial progresses, we think two key problems should be further studied to evaluate the capability of this new data source to infer urban land use. First, the time series model that represents land use type at the BTS level should be improved to enhance urban land use classification. On the one hand, the model should be more sophisticated and incorporate more characteristics (say, the differences between weekdays and between weekends, new indices derived from aggregated mobile phone data) in order to better differentiate between different land use types. This is because the land use is not only dynamically changing, but is often also heterogeneous in some areas. Thus, either the pattern or the volume may not fully interpret the social functions of different land use types. On the other hand, the model should be more transparent to allow an evaluation of the effects of different characteristics on land use classification. This may help us analyze and improve the classification method. Second, because mobile phone data is a new data source in terms of urban planning, it is important to evaluate the uncertainties and influential factors behind land use classification. These include three aspects. One is related to the model, and specifically the different characteristics in the time series. The second concerns the data, particularly the BTS density. The third considers the ground truth, and specifically the heterogeneity of land use.

To overcome these key problems, we construct a new time series by generating a linear combination of the four-day call pattern and volume. This time series not only utilizes more characteristics of mobile phone data, but also makes the classification result easier to interpret. A new semi-supervised scheme is proposed to infer the land use based on this time series. Using this process, we can classify the urban land use and understand the different effects imposed by the call pattern and volume on the classification result. Finally, the uncertainties of land use classification are analyzed in terms of the dissimilarity between land use definition and classification result, mixture of land use, BTS density, and the fuzzy membership value generated by the proposed method.

## 3. Semi-supervised fuzzy c-means (FCM) clustering method for urban land use classification

We first construct a synthesized time series, which is the linear combination of the normalized pattern and the total calling volume. The pattern part can be determined by the characteristics of the mobile phone data that will be used. Then, to determine different types of land use types with the synthetic time series, we use a semi-supervised clustering FCM method. Thus, the effect of different parts of the time series on the classification can be determined by calculating the ratios in the distance between cluster centers and the time series.

The process of classification is divided into the following five steps. 1) Place the aggregated mobile phone data from each BTS into a mesh. 2) Construct the synthesized time series that combines the normalized pattern with the calling volume. A coefficient ( $\beta$ ) is introduced to weight the pattern versus the volume. 3) Determine $\beta$ by training samples of different land uses, which are selected based on expert knowledge. 4) Cluster the time series of mobile phone data using FCM. 5) Post-process the clustering result by assigning each cluster to different land use types. Each of these steps is now described in detail.

### 3.1. Gridding the data

Before being used to identify urban land use, the mobile phone data, aggregated hourly at the BTS level, are interpolated to generate a mesh grid for further computation. The data generated by each cell on an hourly basis form a time series. The procedure is divided into four stages. First, a Voronoi polygon system is generated using the BTS tower locations. Next, the volume in each BTS polygon is divided by its area to give the volume density. The inverse distance weighting (IDW) method is then used to generate the grid at hourly intervals. Finally, the hourly values generated over each BTS form a time series.

### 3.2. Constructing the time series of aggregated mobile data

The time series we use in our method consists of two parts. The first is the hourly pattern of mobile phone data. The second is the total volume, given by:

$$Z_i = [X_i \ \beta \cdot Y_i] \quad (1)$$

, where $Z_i$ $(\{z_{i,j}, i = 1,2,\cdots n; j = 1,2,\cdots T\})$ is the combined time series for cell *i*, $X_i$ $(\{x_{i,j}, i = 1,2,\cdots n; j = 1,2,\cdots T\})$ is the pattern for cell *i* (see equation (2)), *n* is the number of cells in the grid, *T* is the number of hours considered in the pattern, and $Y_i$ is the volume for cell *i* modified by the range transformation (equation (3)).

$$x_{i,j} = \frac{b_{i,j}}{\sum_{j=1}^{T} b_{i,j}} \quad (i = 1,2,\cdots n;\ j = 1,2,\cdots T) \tag{2}$$

$$Y_i = \frac{2[\sum_{j=1}^{T} b_{i,j} - \min(\sum_{j=1}^{T} b_{i,j})]}{\max(\sum_{j=1}^{T} b_{i,j}) - \min(\sum_{j=1}^{T} b_{i,j})} \quad (i = 1,2,\cdots n) \tag{3}$$

, where $b_{i,j}$ is the original hourly calling volume at cell *i*. Note that we multiply the numerator by 2 to ensure that $Y_i$ has the same range as $X_i$. The reason we use range transform is for a comparison of the roles played by the pattern and the volume in the classification.

3.3. Determination of $\beta$

To estimate the coefficient $\beta$, we select $L$ $(L = \sum_1^K l_k)$ samples from *K* land use types ($l_k$ is the number of samples for land use type *k*). These land use types should already be known from other information sources, e.g., points of interest (POI) in Google Earth. The center for each land use sample group ($C_k(\{c_{k,j}, k = 1,2,\cdots K;\ j = 1,2,\cdots T\})$) can be determined by averaging the sample time series:

$$c_{k,j} = \frac{1}{l_k} \sum_{i=1}^{l_k} z_{i,j}^{(k)} \quad (j = 1,2,\cdots T\}) \tag{4}$$

If we define $d_{i,j}$ as the distance between sample *i* and cluster center *j*, then the land use type for sample *i* can be determined by locating the minimum distance between it and each cluster center.

$$ID_i^{'} = \mathrm{find}(d_{i,j} == \min(d_{i,j})) \quad (i = 1,2,\cdots K;\ j = 1,2,\cdots T) \tag{5}$$

$ID'_i$ is the land use type of sample *i*. We define $ID_i$ as the true land use type of sample *i* for the validation. Then the value of $\beta$ can be determined by minimizing the objective function:

$$f(\beta) = \sum_i I(Z_i) \quad (i = 1, 2, \cdots L) \tag{6}$$

, where $I(Z_i) = \begin{cases} 0 & ID'_i = ID_i \\ 1 & ID'_i \neq ID_i \end{cases}$ is an indicator function with $I(\cdot) = 0$ when $Z_i$ is correctly classified; otherwise, $I(\cdot) = 1$. The objective function is calculated for different values of $\beta$. The optimized value of $\beta$ is that at which $f(\beta)$ reaches its minimum.

### 3.4. Determination of final land use type

After determining the value of $\beta$, the time series for all cells are clustered using FCM. There are two strategies to choose the number of clusters in FCM (Bezdek, 1981; Nock and Nielsen, 2006). The first is to simply set the number of clusters to the number of land use types. The second determines the number of clusters from the validation index generated on each execution of FCM (Ray and Turi, 1999). In this study, we choose the second strategy, because certain land use types are the result of a simplified urban planning map, and may thus be a combination of different specific land use types. For example, an Open space may contain areas of Park, Green, Cemetery, and Water. In this context, we would rather retain the natural structure of clusters (which might be some specific land use types) for the post-process combination than generate a predefined number of clusters, which may cause some land use type is divided into different clusters.

### 3.5. Post-processing to assign clusters to specific land use types

Once the clusters have been generated, we perform post-processing to assign each cluster to an appropriate land use type. A cluster is assigned to the specific land use type whose center, as represented by the samples used in section 3.3, is closest to the center of the cluster. If the number of clusters is greater than the number of land use types, at least one land use type will be assigned more than one cluster. If there are fewer clusters than land use types, then we use the number of land use types to re-cluster the data.

## 4. Aggregated mobile phone data from Singapore

The mobile phone data used for the land use classification are the hourly aggregated number of calls managed by each of 5500+ BTS towers in Singapore. To determine land use types from mobile phone data, we use data from a whole week (Monday 28 March to Sunday 3 April, 2011). Based on the timelines of mobile phone data for these seven days, we use the linear combination of the normalized pattern and the call volume. The pattern is a four-day mode, i.e., general weekday, Friday, Saturday, and Sunday, where the general weekday is the average pattern for Monday, Tuesday, Wednesday, and Thursday. To clarify our choice of the four-day mode, we consider the normalized timeline (i.e., the pattern) between different days (Table 1). We choose the four-day mode for two reasons. First, Monday, Tuesday, Wednesday, and Thursday are similar, and can be considered as one mode. From Table 1, we can see that the three closest neighbors to each of Monday, Tuesday, Wednesday, and Thursday are all from these four days themselves. For example, Tuesday, Wednesday, and Thursday are closer to Monday than the other three days (i.e., Friday, Saturday, and Sunday) in terms of the normalized pattern distance. (Interestingly, in most cases, the temporally closer are any two of these four days, the smaller the time series distance between them.) Therefore, the data for Monday–Thursday are averaged to represent an ordinary weekday. Second, Friday, Saturday, and Sunday show significant differences, and can be considered as three separate modes. Table 1 indicates that each of Friday, Saturday, and Sunday are far away from all the other days. As a result, we choose this four-day mode for land use classification. This ordinary weekday and the remaining three days form a 96-point time series. The comparison of the detection rate between the four-day mode, the two-day mode (an average weekday and an average weekend) and the seven-day mode also confirms that this processing generates the best classification result (see the discussion in the supplementary document).

Table 1. Distance of normalized pattern between different days

| | Mon. | Tue | Wed | Thu | Fri | Sat | Sun |
|---|---|---|---|---|---|---|---|
| Mon | 0 | 0.0049 | 0.0089 | 0.0103 | 0.0175 | 0.0245 | 0.0388 |
| Tue | 0.0049 | 0 | 0.0057 | 0.0072 | 0.0137 | 0.0224 | 0.0359 |
| Wed | 0.0089 | 0.0057 | 0 | 0.0067 | 0.0099 | 0.0223 | 0.0332 |
| Thu | 0.0103 | 0.0072 | 0.0067 | 0 | 0.0113 | 0.0201 | 0.0301 |
| Fri | 0.0175 | 0.0137 | 0.0099 | 0.0113 | 0 | 0.0216 | 0.0283 |
| Sat | 0.0245 | 0.0224 | 0.0223 | 0.0201 | 0.0216 | 0 | 0.0231 |
| Sun | 0.0388 | 0.0359 | 0.0332 | 0.0301 | 0.0283 | 0.0231 | 0 |

In order to validate the clustering result, we use the urban planning map of Singapore, taken from the website http://www.ura.gov.sg/uramaps/?config=config_preopen.xml&preopen=Master%20Plan, and combine land use types to form the ultimate map (Figure 1). Here, we have divided Singapore into five land use types: Residential, Business, Commercial, Open space, and Others. Prior to classification, we interpolate the aggregated hourly data into a 200 m × 200 m grid using IDW, and generate 96 pattern layers and one volume layer.

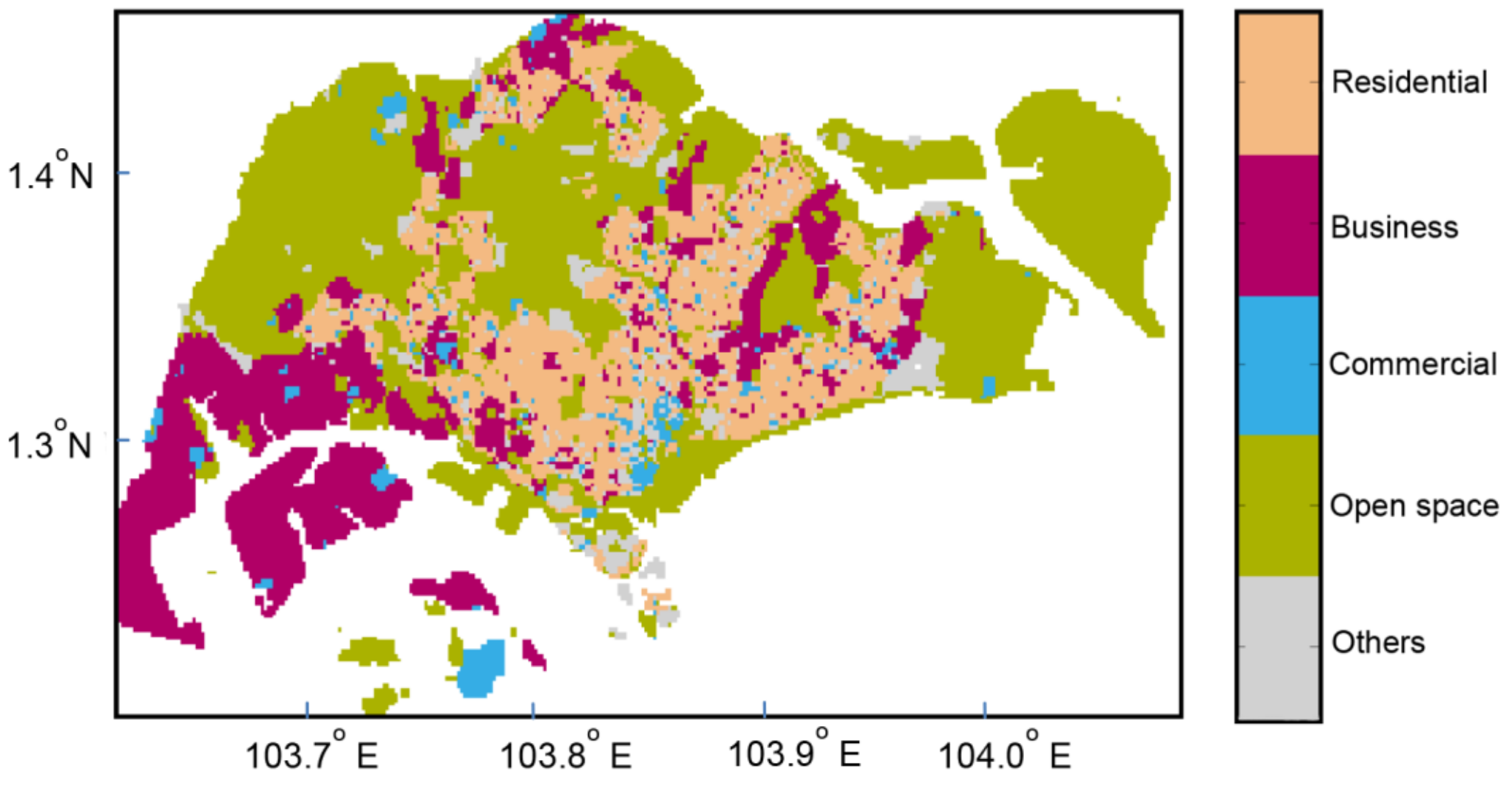

Figure 1. Land use in Singapore

## 5. Land use classification for Singapore

5.1. Determination of land use types

After generating 97 image layers, the first 96 are transformed using equation (2) to generate $X_i$, and the final layer is transformed using equation (3) to generate $Y_i$. As discussed above, we combine the pattern ($X_i$) and the volume ($Y_i$) to form a new time series $Z_i$ using the coefficient $\beta$ (see equation (1)). Next, we determine the value of $\beta$ through the following training process. First, 105 samples (allocated based on the prior knowledge of the areas of different land use types: 25 samples each for Residential, Business, and Open space, 20 samples for Commercial, and 10 samples for Others) are chosen based on remote sensing imagery and POI data (from Google Earth) as well as information provided by several residents of Singapore. To ensure the samples represent their land use types, we select them according to three criteria. First, samples are picked from homogeneous areas. Second, we avoid samples from near the boundary between different land use types. Third, we attempt to pick samples that are close to a BTS tower. The objective function $f(\beta)$ is calculated at different values of $\beta$, and the results are shown in Figure 2. We can see that the minimum value is acquired when $\beta$ is between 0.65 and 0.80.

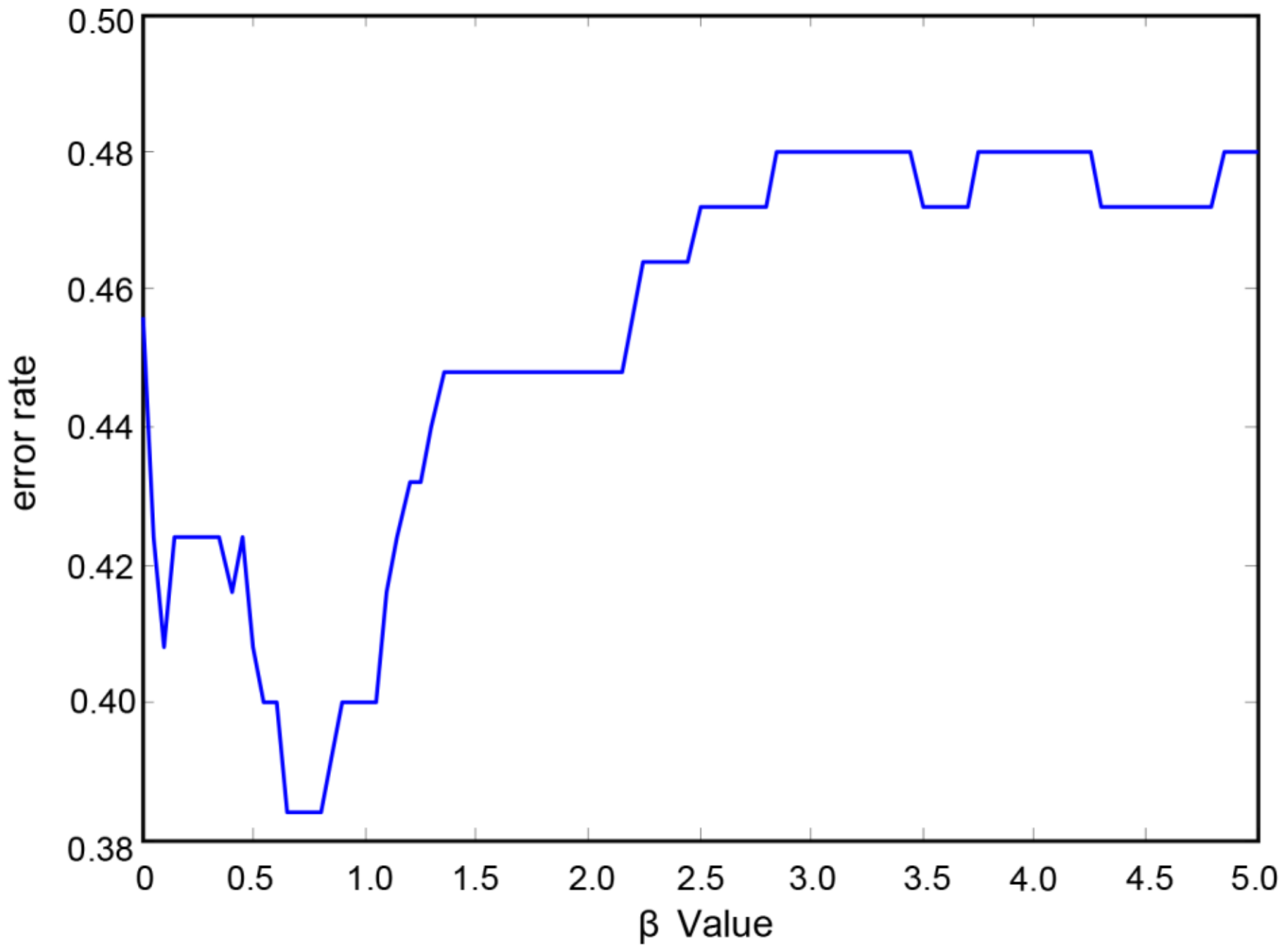

Figure 2. Error rates generated at different values of $\beta$

The sample centers of different land use types are shown in Figures 3 and 4. Figure 3 shows the pattern part of the centers, each of which contains 96 points. Figure 4 is a boxplot of the volume of each land use. We can see that all land use types can be characterized by a combination of pattern and volume. For example, Residential areas are characterized by a similar size pattern for each of the four days and medium volume, whereas Business areas are characterized by a high-thin pattern on the ordinary weekday and Friday, a low weekend pattern, and low volume. The other land use types can be similarly characterized. The characteristics of each time series guarantee the classification of land use type.

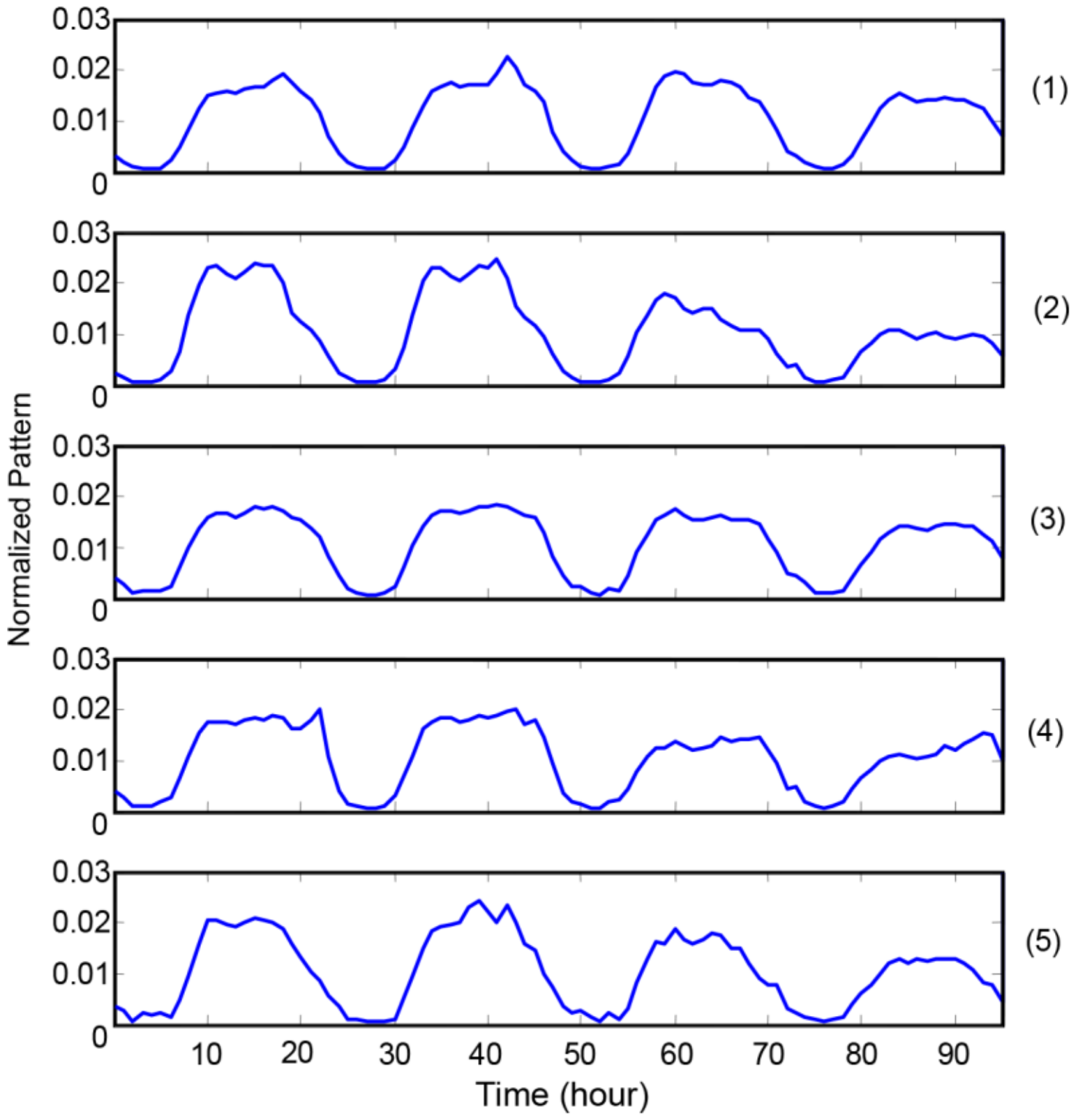

Figure 3. Patterns of centers of time series samples with $\beta = 0.75$

(1-Residential; 2-Business; 3-Commercial; 4-Open space; 5-Others)

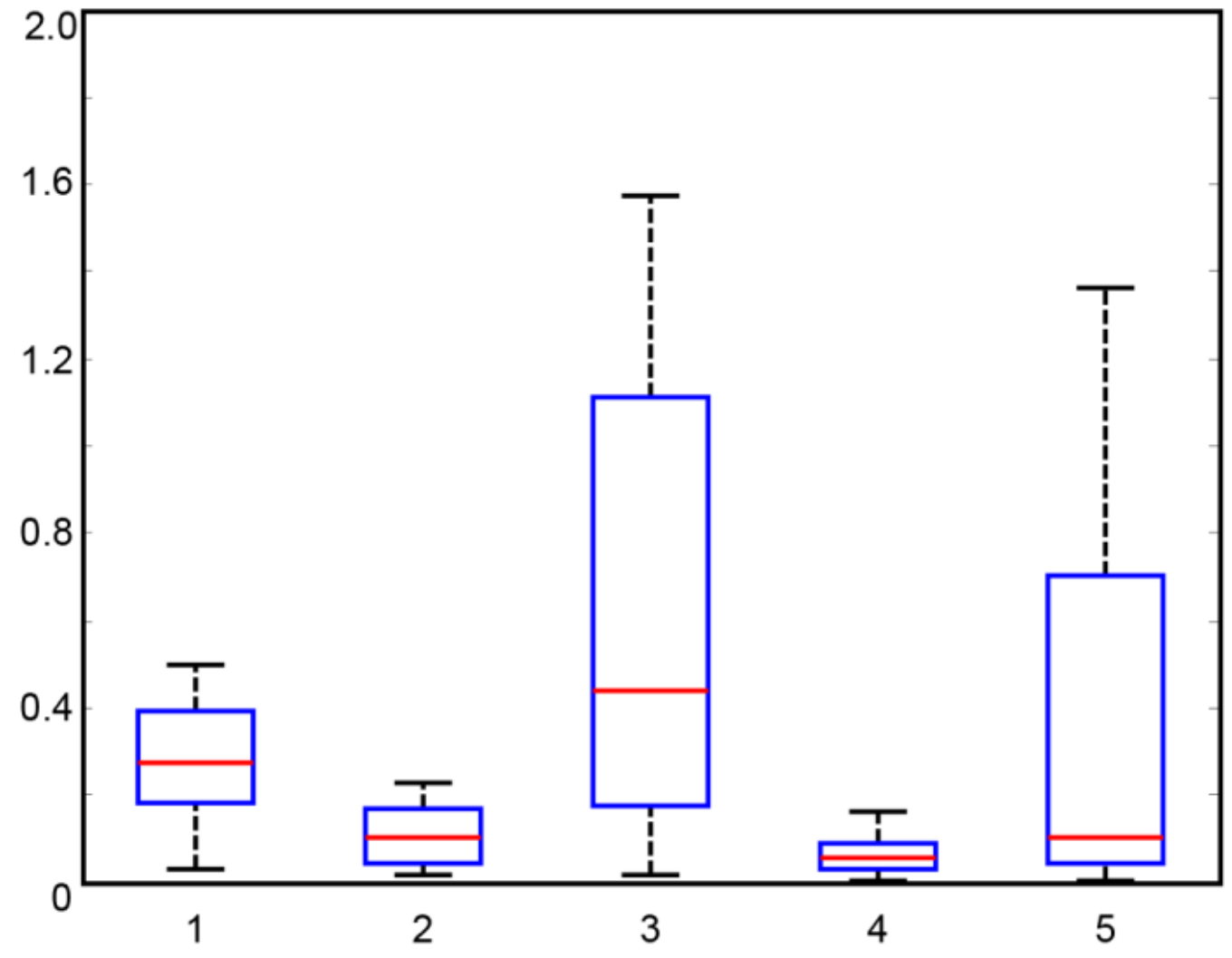

Figure 4. Volume of time series samples with $\beta = 0.75$

(1-Residential; 2-Business; 3-Commercial; 4-Open space; 5-Others)

## 5.2. Clustering result

We use FCM to cluster the aggregated data by setting $\beta$ to 0.75, based on the training result. The cluster number is determined by the validity indices, which indicate that the optimum cluster number is 6. After post-processing, two clusters are combined and determined as Open space. Finally, we generate the land use map displayed in Figure 5(a).

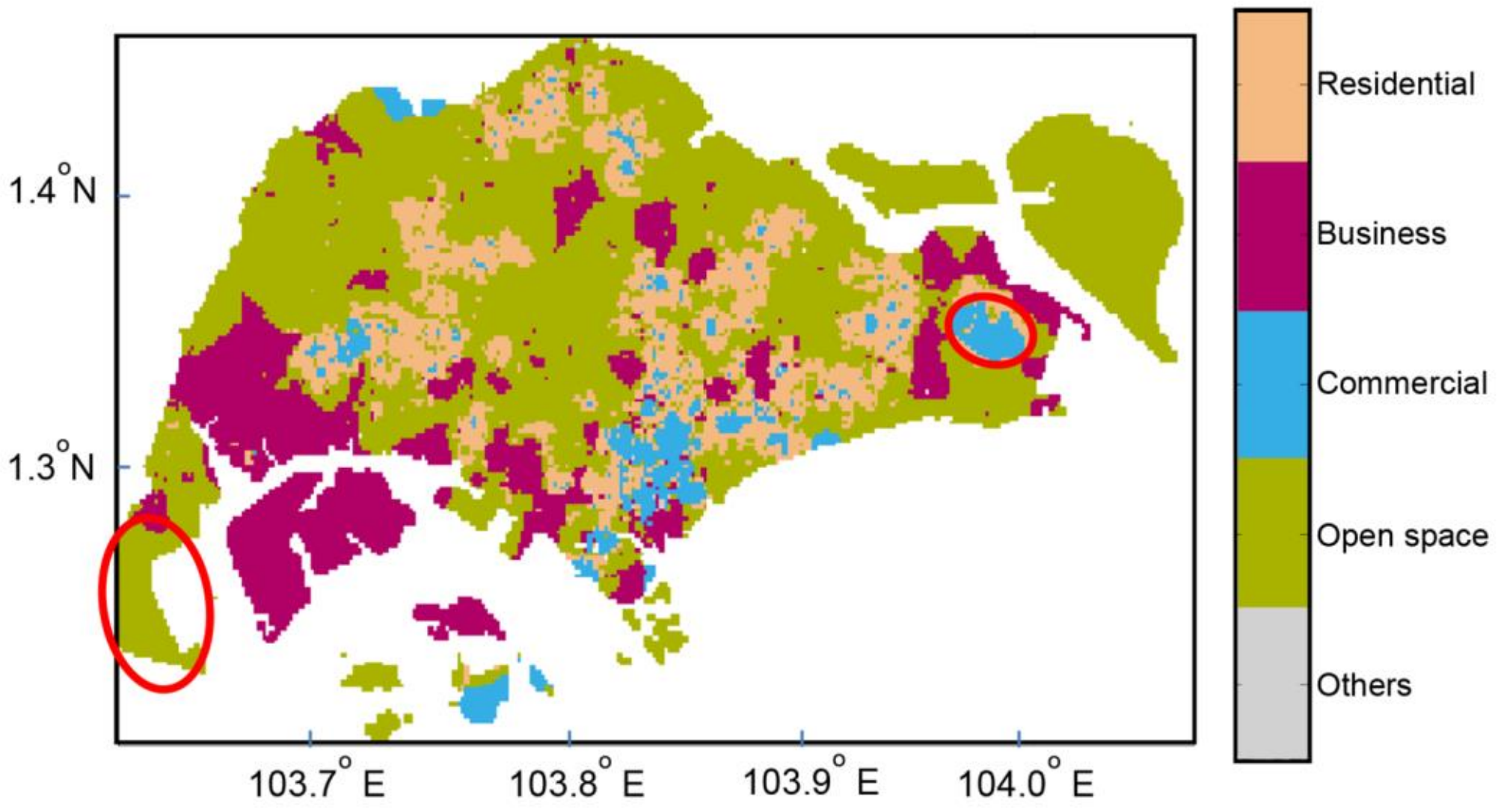

(a)

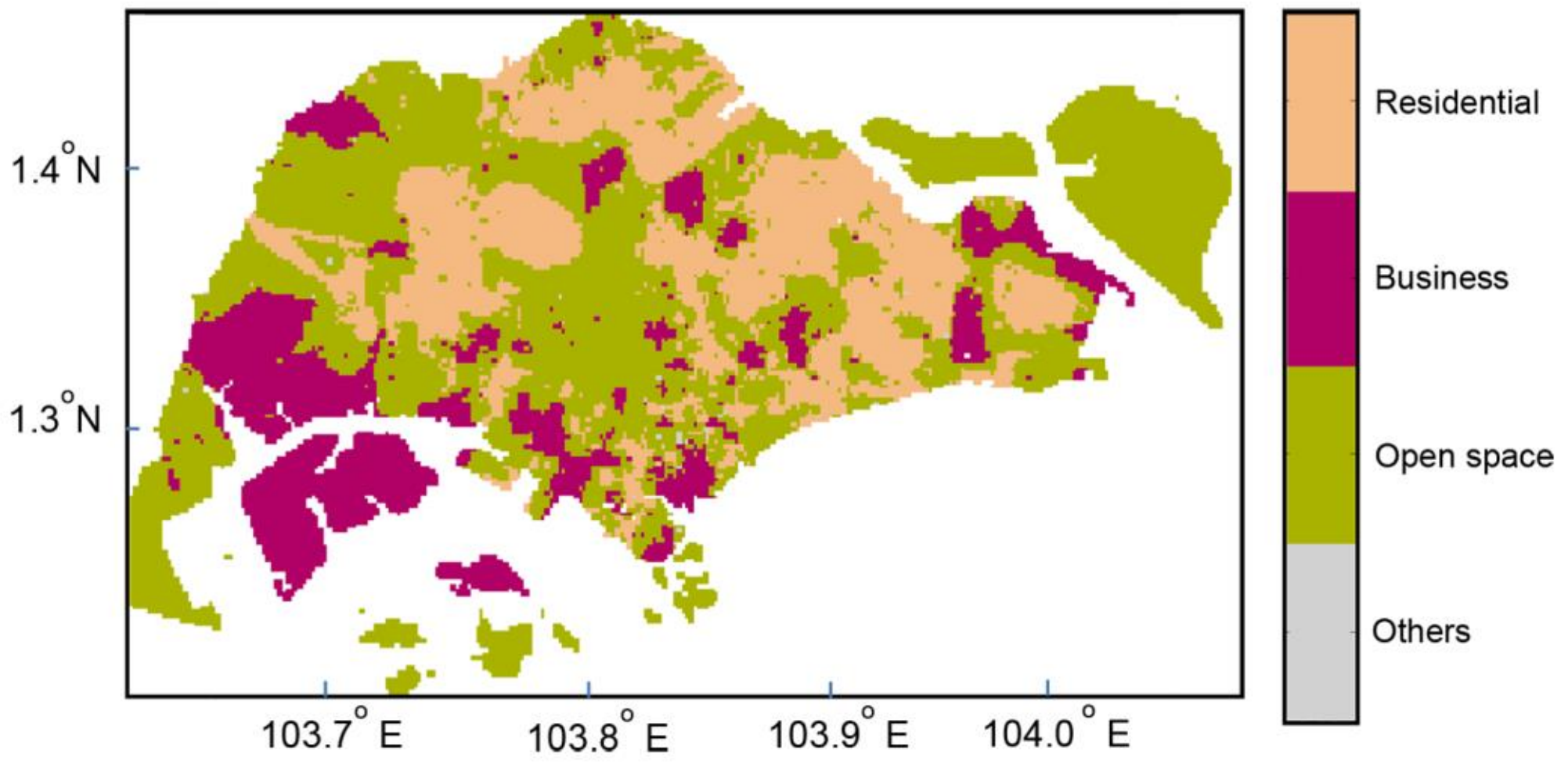

(b)

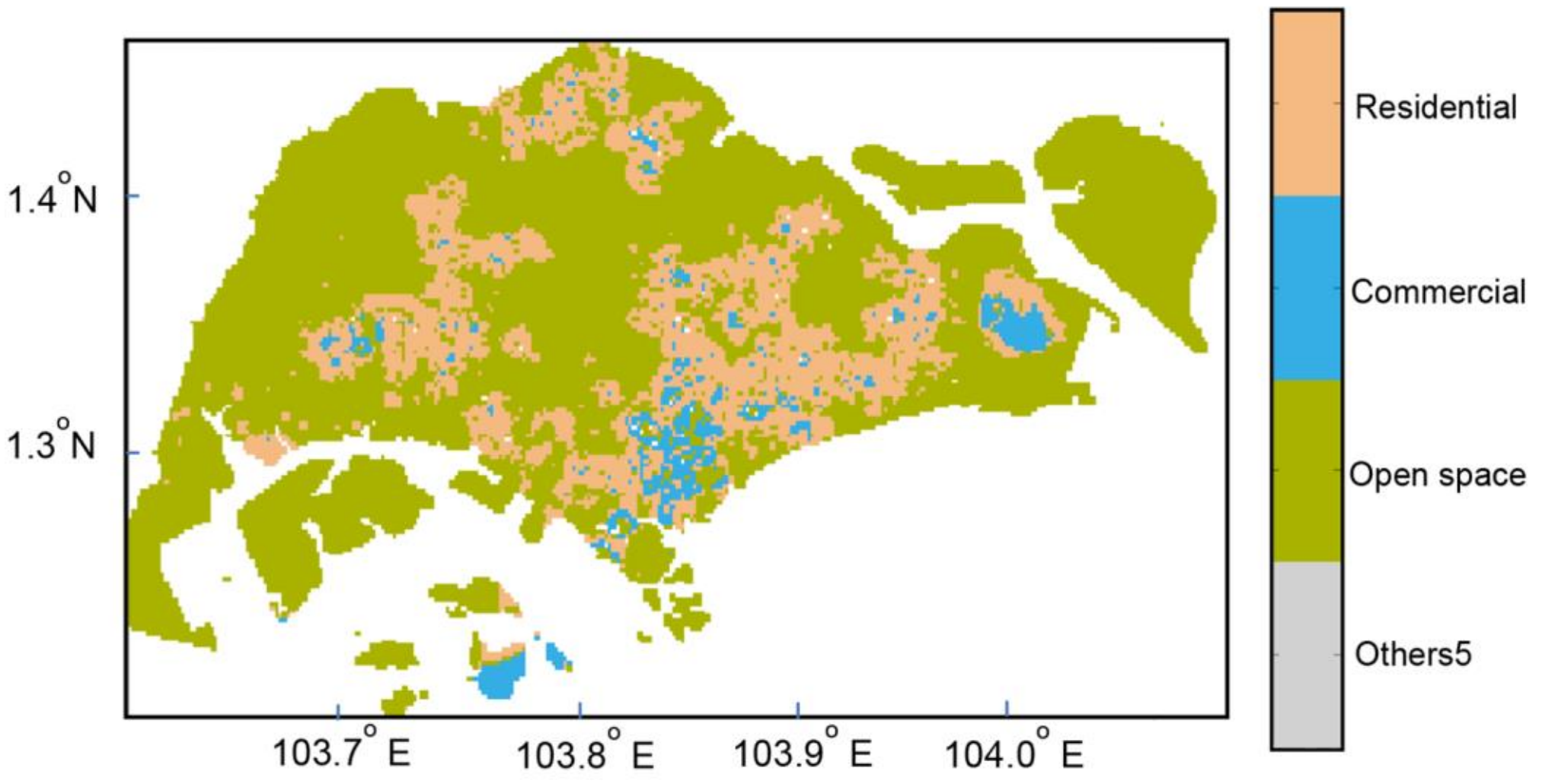

(c)

Figure 5. Clustering result for land use types in Singapore

(a) Classification generated from the synthetic time series (detection rate: 58.03%; the left red ellipse indicates the area defined as Commercial in Figure 1 is identified as Open Space; the right red ellipse indicates the area defined as Open Space in Figure 1 is identified as Commercial). (b) Classification generated from the pattern data (detection rate: 52.58%). (c) Classification generated from the volume data (detection rate: 52.68 %).

Comparing the classification result with the urban planning map (Figure 1), we find that all land use types are identified with an overall detection rate of 58.03%, which is close to that generated by Toole et al. (2012) (The detection rate is 54%). In the supplementary document, we also showed that four-day mode generates the highest detection rate compared with that for two-day mode (57.65%) and for seven-day mode (55.15%). The confusion matrix is shown in Table 2. From this table, we can see that the order in which the land use types are best detected is Open space,

Residential, Business, Commercial, and Others (this can be determined from the diagonal elements in the matrix, which mean the land use is correctly classified). Only Residential, Business, and Open space land use types have rates close to or above 50%. The detection rates of Commercial and Others are less than 50%. In addition, some land use types have a misclassification rate of over 30%. Overall, land use is most commonly misclassified as Open space, while Others is the most likely to be misclassified.

Table 2. Confusion matrix of the classification

| | Residential | Business | Commercial | Open space | Others |
|---|---|---|---|---|---|
| Residential | **0.4912** | 0.0490 | 0.0658 | 0.3938 | 0.0002 |
| Business | 0.0978 | **0.5018** | 0.0174 | 0.3825 | 0.0005 |
| Commercial | 0.1612 | 0.1535 | **0.3457** | 0.3302 | 0.0093 |
| Open space | 0.0769 | 0.1210 | 0.0395 | **0.7622** | 0.0004 |
| Others | 0.0037 | 0.1737 | 0.0772 | 0.5026 | **0.2428** |

To determine the reasons for this particular land use classification, we draw the center of each real land use type and that of each cluster in Figure 6. Comparing the two, we find that the Residential, Business, and Open space regions generated by our method show both a similar pattern (Figure 6a and c) and volume (Figure 6b and d) as the real land use types. Although Others in Figure 6a shows a similar pattern to the real one ("5" in Figure 6c), its volume ("5" in Figure 6b) is somewhat different (Figure 6d). The Commercial volume ("3" in Figure 6b) suggested by the clustering has a larger value than the actual volume (“3” in Figure 6d), and its pattern is also different (“3” in Figure 6a and c). This shows why Residential, Business, and Open space have high detection rates while Commercial and Others have lower ones.

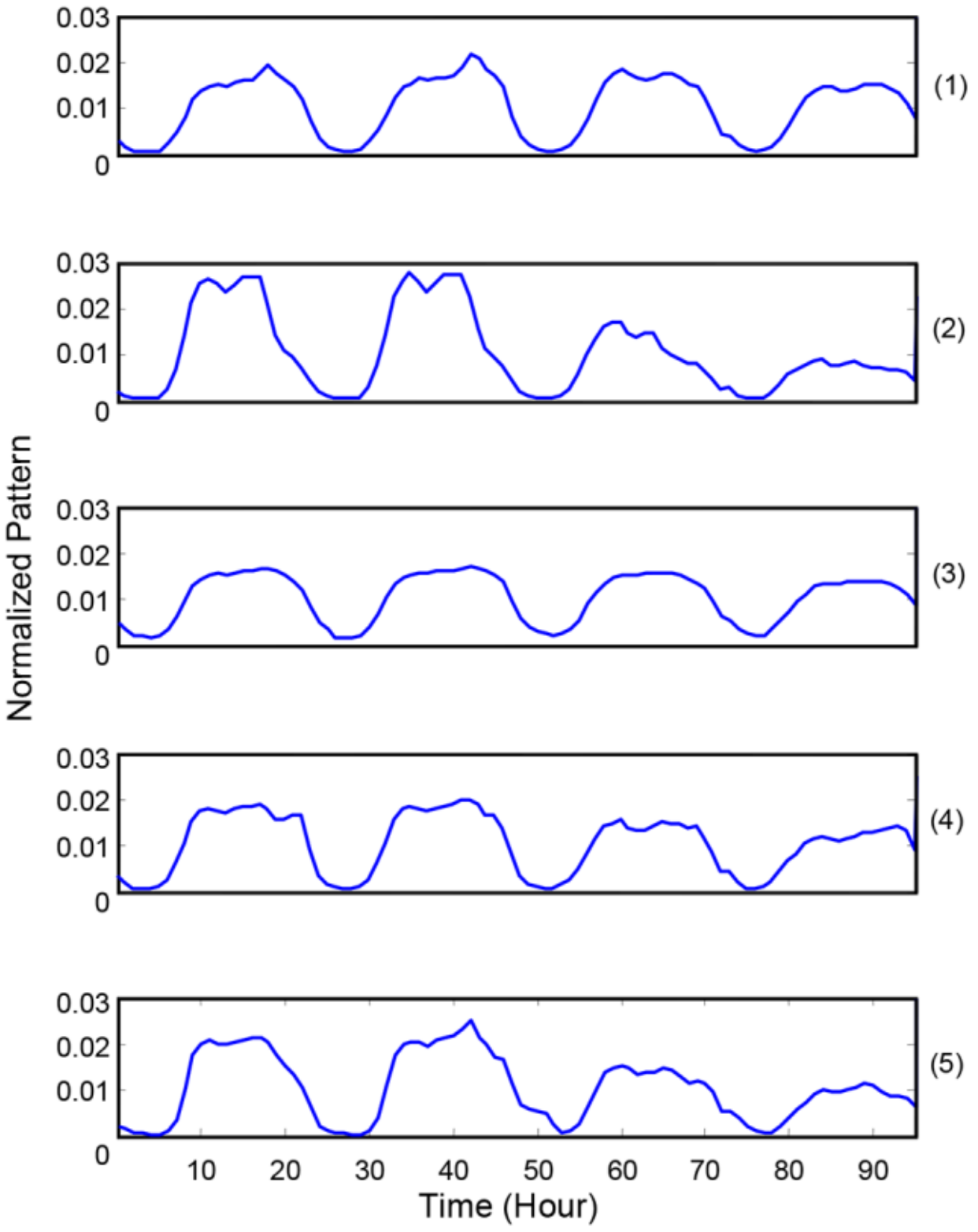

(a)

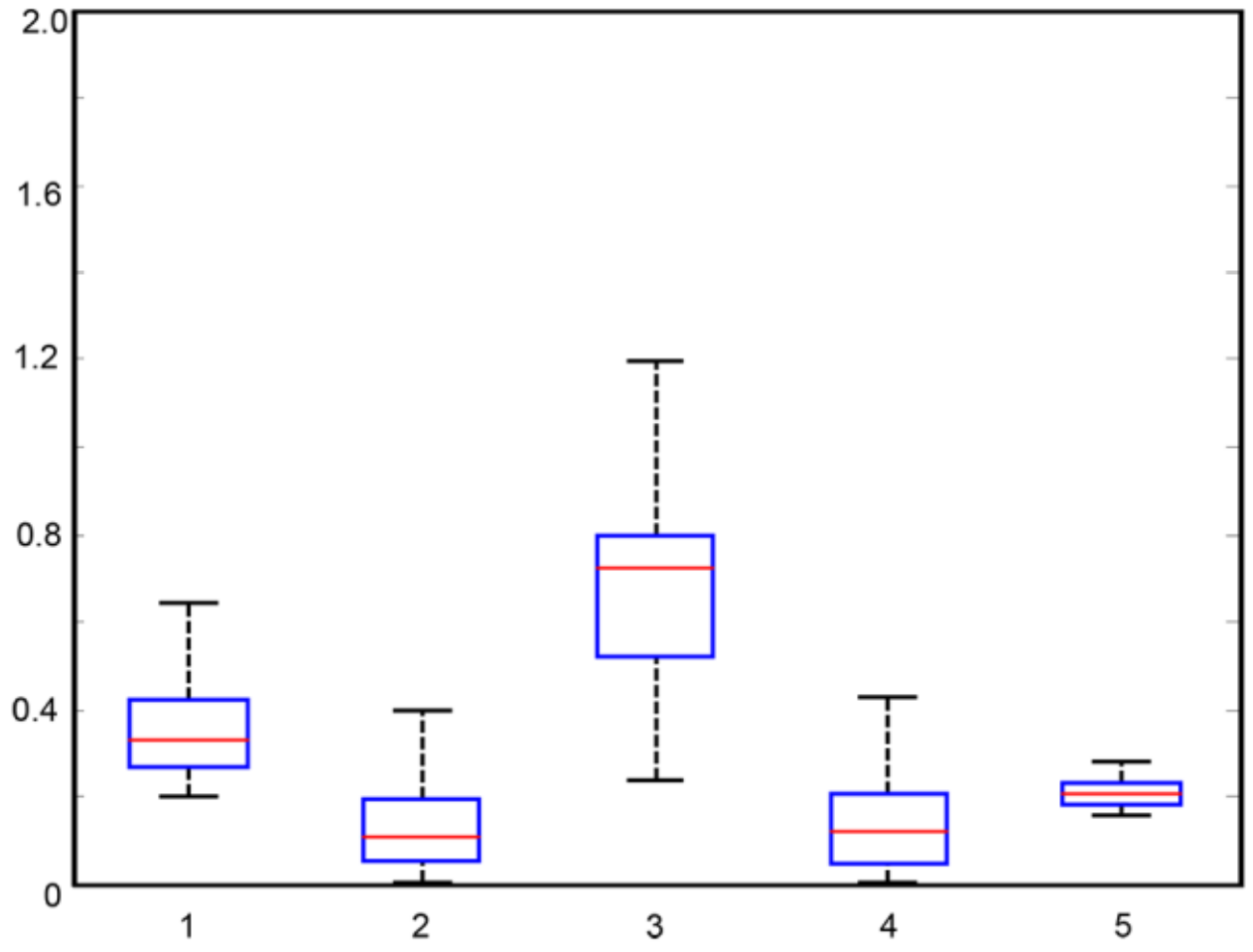

(b)

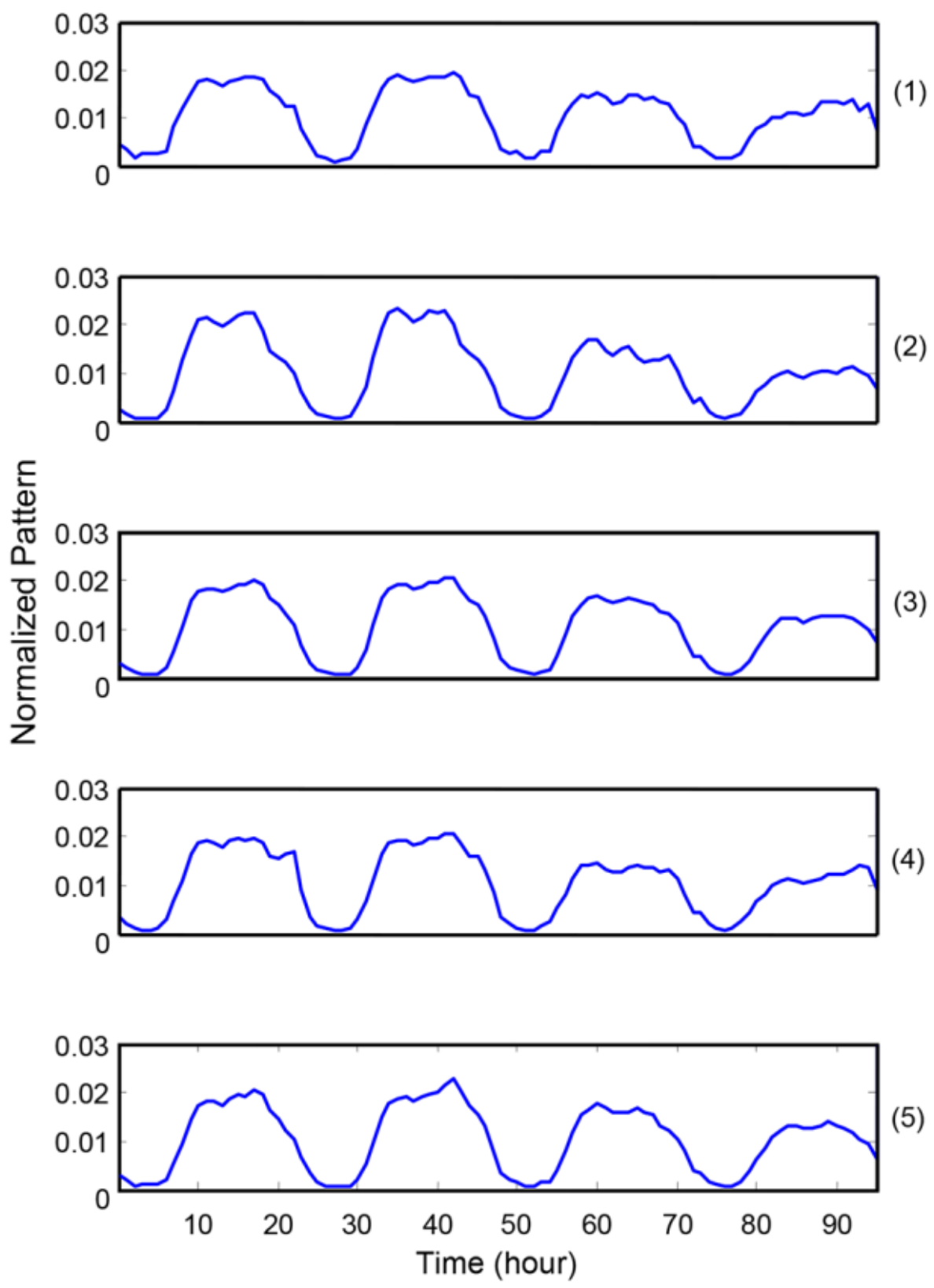

(c)

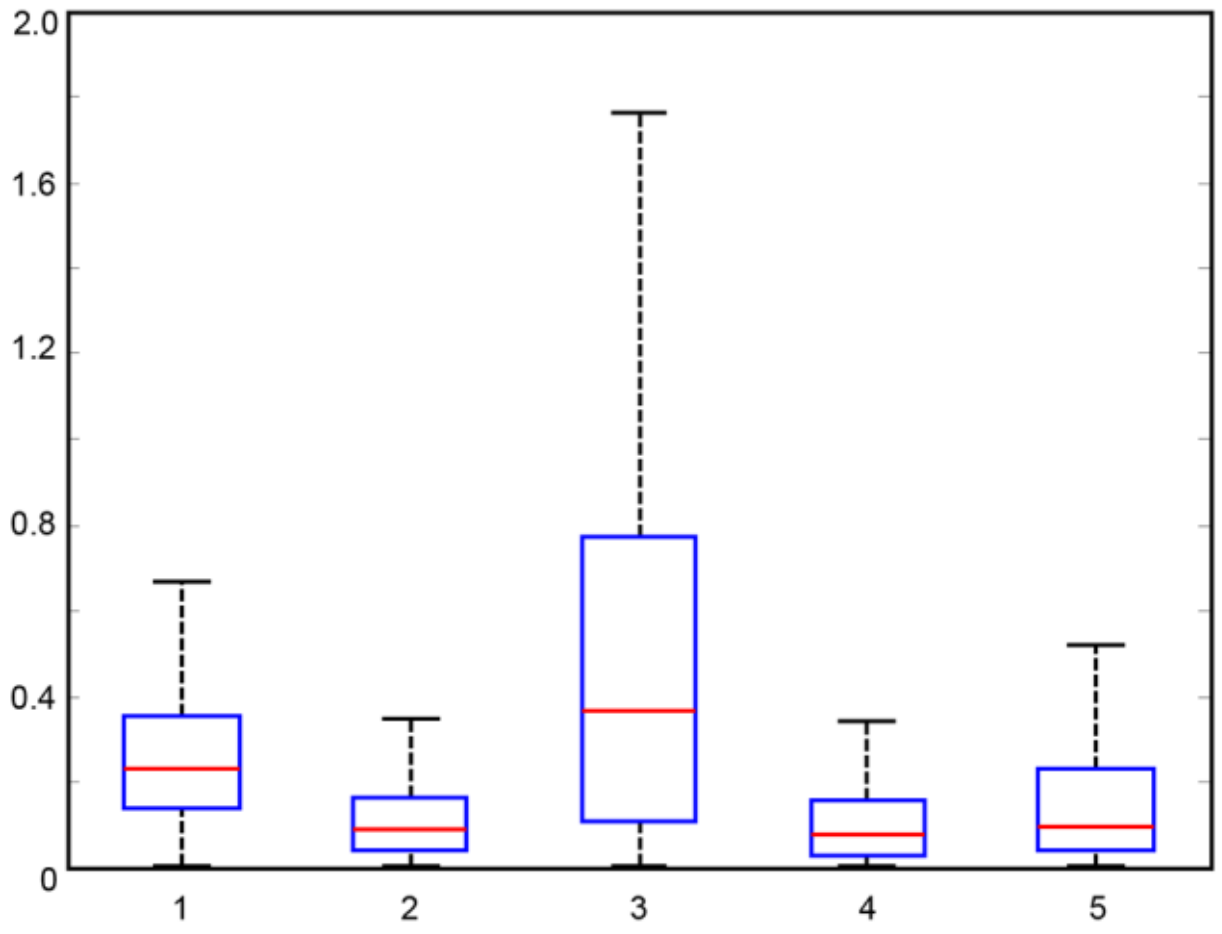

(d)

Figure 6. Centers of clusters for different land use types

(1-Residential; 2-Business; 3-Commercial; 4-Open space; 5-Others)

(a) Centers of pattern for classification; (b) Distribution of volume for classification;

(c) Centers of patterns for known land use; (d) Distribution of volume of known land

use

## 5.3. Evaluation of the effect of call pattern and volume on classification

We now examine how the value of $\beta$ influences the detection rate. The detection rate calculated for different values of $\beta$ is shown in Table 3. The detection rate generally increases with $\beta$ until $\beta = 0.75$, then decreases for $\beta$ > 0.75.

Table 3. Change in detection rate with $\beta$ (four-day mode)

| $\beta$ value | 0 | 0.15 | 0.30 | 0.40 | 0.50 | 0.60 | 0.65 | 0.70 |
|---|---|---|---|---|---|---|---|---|
| Detection rate (%) | 52.58 | 54.30 | 55.12 | 57.56 | 56.50 | 57.51 | 57.57 | 57.97 |
| $\beta$ value | 0.75 | 0.8 | 0.9 | 1.00 | 1.25 | 1.50 | 2.50 | ∞ |
| Detection rate (%) | 58.03 | 57.30 | 55.61 | 55.44 | 54.54 | 54.24 | 54 .01 | 52.68 |

As discussed in Section 2, the distance between samples and the cluster centers is calculated during the FCM algorithm. The distance consists of two parts. The first ($d_1$) is the distance between the patterns, and the second ($d_2$) is that between the volumes weighted by $\beta$. Essentially, the value of $\beta$ represents the balance between call pattern and call volume, both of which are normalized. As $\beta$ decreases, the weight of the pattern part in the overall distance between samples and centers will increase. On the contrary, as $\beta$ increases, the weight of the volume part will increase. The next issue is to determine which part dominates the distance (i.e., the difference in discerning between land use types) in the classification generated at the optimized value of $\beta$ ($\beta = 0.75$). We calculated the ratio between $d_1$ and $d_2$ for all land use types classified with $\beta = 0.75$. The results are given in Table 4. From this table, we can see that the ratio is greater than 1 for all land use types except Commercial. The average ratio is 1.6471, which indicates that the distance between the patterns is generally larger than those between the weighted volumes. The ratios for different land use types implies that the pattern information plays a more important role in the classification for all land use types, with the exception of Commercial areas. This is also consistent with the differences in the time series of different land use types, which can be found in Figure 6. Specifically, Commercial has the highest volume, which is significantly different from the other land use types. This causes the volume to play a more important role in separating Commercial from the other types. On the contrary, the other land use types show more significant differences between the patterns than the volume, which leads to the larger distances between the patterns. This analysis of the effect of the call pattern and volume shows that our method can utilize different characteristics of mobile phone data to differentiate between land use types.

Table 4. Ratio between pattern and volume for different land use types

| Land use type | Residential | Business | Commercial | Open space | Others | Average |
|---|---|---|---|---|---|---|
| Ratio between Pattern and volume | 1.1462 | 2.0758 | 0.9594 | 2.5467 | 1.5072 | 1.6471 |

## 6. Comparison between classifications using different information

To further validate the method based on the newly constructed time series, we compare the classification with that generated with either the pattern or the volume. The clustering validity index shows that five clusters are generated for pattern information only, while four clusters are generated for the volume. The results are shown in Figure 5b and c. Figure 5b indicates that the clustering based on the pattern information did not identify Commercial areas, and Figure 5c indicates that the clustering based on volume data did not identify the Business regions. The overall detection rates are also lower (52.58% for pattern and 52.68% for volume) than that based on the combination of pattern and volume.

The pattern information fails to identify Commercial areas because these are highly mixed with Residential areas. According to the Master Plan 2008 of Singapore, more than 45% of the Commercial area is either "residential with commercial on the first floor" or a "mixture of commercial and residential". This highly mixed distribution causes difficulties in discerning Residential from Commercial. To quantify the degree of mixing between different land use types, we can calculate the posterior classification based on the pattern information, in which the land use type over a cell is determined by locating the minimum distance between the pattern part and the centers of known land use types. We generate the posterior confusion matrix by comparing the posterior classification with the Master Plan 2008 (Table 5). This shows that only 9.89% of Commercial areas are correctly classified, with 40.54% mixed into Residential. This also explains why the Commercial land use type is not identified from pattern information alone.

Table 5. Posterior confusion matrix of pattern information

| | Residential | Business | Commercial | Open space | Others |
|---|---|---|---|---|---|
| **Residence** | 0.6708 | 0.0731 | 0.0571 | 0.0138 | 0.1852 |
| **Business** | 0.1299 | 0.5842 | 0.0279 | 0.2285 | 0.0296 |
| **Commercial** | 0.4054 | 0.2679 | 0.0989 | 0.1032 | 0.1246 |
| **Open space** | 0.1645 | 0.3297 | 0.0557 | 0.3478 | 0.1024 |
| **Others** | 0.4640 | 0.2685 | 0.0462 | 0.0483 | 0.1729 |

The classification based on volume fails to detect Business land use because this volume shows no significant difference from that of Open space. The box plot of each land use type is shown in Figure 6d, indicating that Business ("2" in the figure) and Open space ("4" in the figure) have very similar median values and ranges. In this case, these two land use types cannot be separated merely by their volume, which cause only four land use types to be identified.

## 7. Discussion

In this section, we analyze the possible causes of errors generated by our method. There are four factors that may affect the error rate of the classification. The first is the difference between the definition of land use in urban planning and the function derived from the mobile phone data. The second is the degree of heterogeneity of different land use types (i.e., different land use types are mixed in the same area). The third is the precision of the information recorded, which is related to the density of BTSs in each cell. The fourth is the fuzzy membership threshold ($\alpha$ -cut) used in FCM.

7.1. Dissimilarity between definition of land use and that derived from the mobile phone data

Previous research has found that zoned areas are not necessarily used as intended, which may lead to incorrect classification (Soto and Frias-Martinez, 2011a; Toole et al., 2012). However, these studies only provided some examples, without summarizing all scenarios. Here, we try to list all possible situations. The first is when various social activities are conducted on one land use type. As mentioned above, a large portion of the residential area in Singapore is mixed with the commercial area. The second is the heterogeneity of a land use type. For example, the airport is a homogenous area in the Master Plan 2008, but the landing area and the terminals in the airport are different in terms of social function. Thus, in the result generated by the mobile phone data, the terminal is classified as Commercial, whereas the landing area is classified as Open space (Figure 5a). This is because the terminal exhibits a very high volume, while that of the landing area is very low. The third is that some areas with specific uses are reserved for other uses in the future. For example, the western part of the business area located in southwest Singapore is “misclassified” as Open space by the mobile phone data (Figure 5a). In fact, this area is an empty space (this can be confirmed from remote sensing images in Google Earth) that is reserved for future business use.

7.2. Correlation between the error rate and BTS density

As we know, the volume of each BTS is calculated by aggregating the number of calls in the polygon generated by Voronoi tessellation (Okabe et al., 2000). When the BTS density is low (i.e., the area of the Voronoi polygon is large), there is a risk that the volume may include calls from areas of different land use. On the contrary, when the BTS density is high, calls collected in this area will have less “interference”, i.e., the signal is “purer”. In order to determine if the purity of signal affects the precision of land use classification, we calculated the detection rates for different BTS densities (Table 6). Note the density in this table is represented by the number of BTSs in each cell. From the table, we can see that the detection rate increases with the BTS density, except when the density is 0. Interestingly, the detection rate attains a relatively high value (i.e., 60.56%) when the density is 0. This is because most of the cells that have a density of 0 are Open space. As the signals in Open space are “purer”, the detection rate in these cells is high. As a result, we can conclude that the “purer” the signal recorded by a BTS (either in the homogenous and large areas with low BTS density or in areas with a high BTS density), the higher the precision of the classification.

Table 6. Relationship between error rate and BTS density

| Towers Density | 0 | 1 | 2 | 3 | 4 | 5 | 6 | 7 | 8 | 11 |
|---|---|---|---|---|---|---|---|---|---|---|
| Detection rate (%) | 60.56 | 44.81 | 50.78 | 51.18 | 52.94 | 57.14 | 58.82 | 75.00 | 75.00 | 100.00 |
| Number of cells | 16548 | 2522 | 963 | 211 | 68 | 21 | 17 | 4 | 4 | 1 |

7.3. Relationship between error rate and mixture entropy

Another factor that might influence the precision is the mixture of the land use. Because the resolution of Singapore's Master Plan 2008 is much higher (4 m) than that of our classification (200 m), we can calculate the error rates in terms of the land use entropy ($En_j$), which measures the randomness of the areas of different land use types in each cell as:

$$\mathrm{En}_j = -\sum_i \mathrm{p}_{i,j} \ln(\mathrm{p}_{i,j}) \quad (7)$$

, where $\mathrm{p}_{i,j}$ is the occupancy rate of the area of land use type *i* in cell *j*.

The relationship between the error rate and the land use entropy is shown in Figure 7. It is interesting to see that the error rate increases with the land use entropy. The reason for this is obvious. If the entropy of a cell is high, which means more land use types coexist in the cell (i.e. the cell is more heterogeneous), then the error rate of the classification increases. The average entropy for residential, business, commercial, open space and others are 0.42, 0.18, 0.47, 0.084 and 0.57, respectively. We can see that the lower the entropy of some land use type, the higher the detection rate (Table 2).

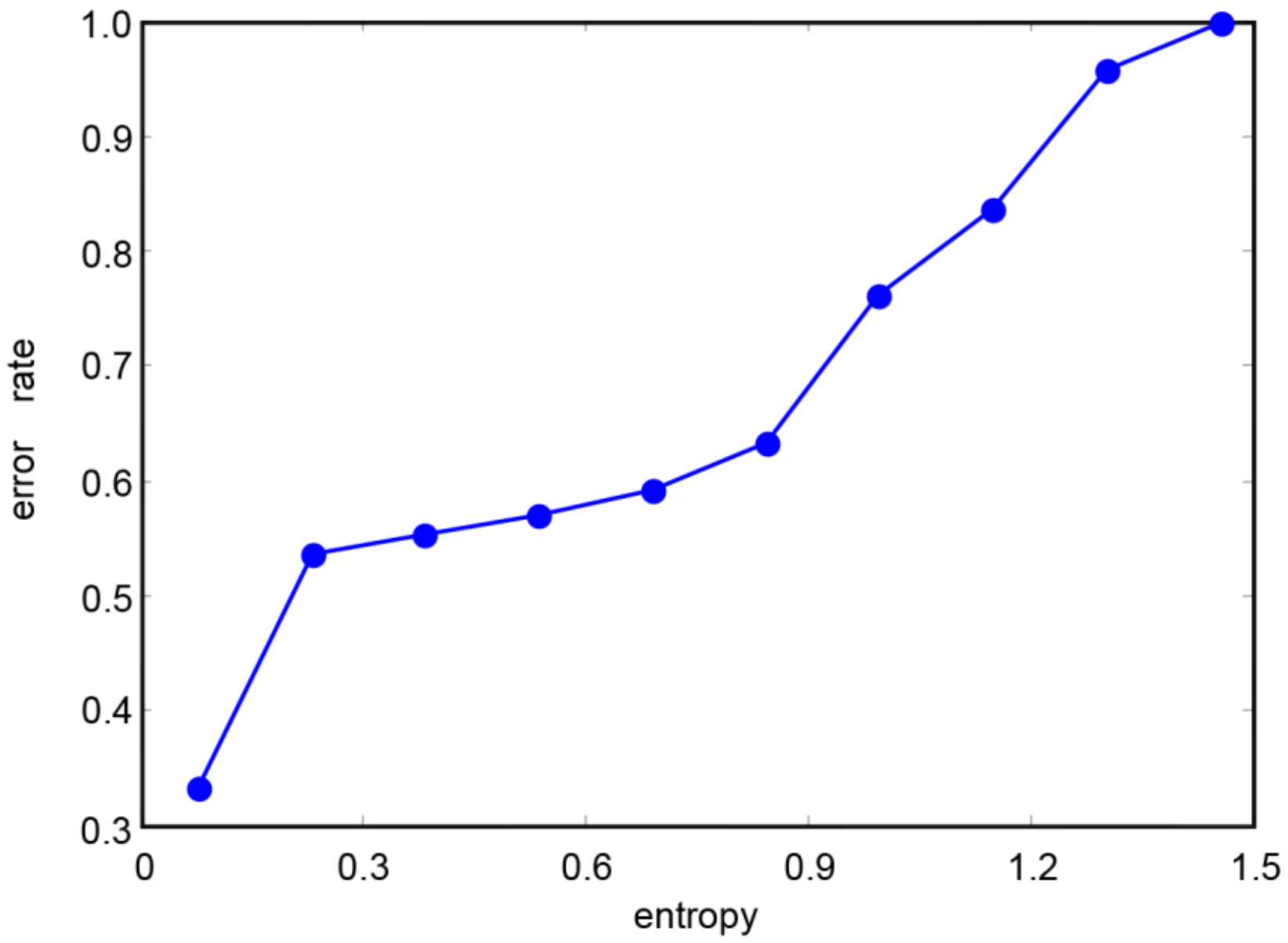

Figure 7. Relationship between land use entropy and error rate

7.4. Relationship between error rate and fuzzy membership value

As we know, the FCM result includes the fuzzy membership value of a sample belonging to each cluster for a certain value of $\alpha$ -cut. Our question is: how will the detection rate change if we change the value of $\alpha$ -cut? The detection rates obtained with different $\alpha$ -cut values are listed in Table 7. We can see that the detection rate is 60.39% when $\alpha$ -cut is 0.5, and that 85.46% of the total area has a membership value greater than 0.5. As α-cut increases to 0.8, only 45.32% of the total area attains this membership value, although the detection rate increases to 72.89%. We can conclude that the detection rate increases with $\alpha$ -cut, but must bear in mind that the area with such a detection rate will decrease.

Table 7. Detection rates at different values of $\alpha$ -cut

| **Value of $\alpha$-cut** | **0.5** | **0.6** | **0.7** | **0.8** | **0.9** |
|---|---|---|---|---|---|
| Detection rate (%) | 60.39 | 61.10 | 65.41 | 72.89 | 88.73 |
| Percentage of area with membership value larger than $\alpha$-cut | 85.46 | 73.35 | 60.27 | 45.32 | 29.16 |

## 8. Conclusions and future work

In this paper, we constructed a synthesized time series of mobile phone activity to identify land use types using a semi-supervised clustering method. The synthesized time series was obtained as a linear combination of the (four-day) pattern and the volume of aggregated data by introducing the weighting coefficient $\beta$. Our classification of land use in Singapore produced a detection rate of 58.03% with $\beta$ set to its optimized value of 0.75, as determined by a training process. Comparisons show that: (1) the data combining both the pattern and volume generate better classifications than those based on either the pattern or the volume alone; (2) four-day mode generates the higher detection rate than that of two-day mode and that of seven-day mode. We can analyze the importance of different parts of the constructed time series on the overall classification, as well as on each type of land use. The results show the relative importance of 'pattern' over 'volume' in detecting most land use types.

We also determined some factors that influence the accuracy of the land use classification. First, there are substantial differences between the urban planning map and the land use retrieved from mobile phone data. Second, areas of mixed land use result in heterogeneous mobile phone usage, and thereby increase the error rate. Third, the purity of the signal in each cell, essentially the BTS density, influences the precision of classification. In general, the higher the density, the higher the precision generated by the classification, except for areas where the density is 0. This indicates that land use classification based on mobile phone data might generate good results in areas with a high BTS density and pure land use types.

Our analysis shows that mobile phone data can reveal the social function of land use. Nevertheless, the overall detection rate of less than 60% indicates that mobile phone data alone are not adequate for urban land use classification, although in some areas the data generate relatively high detection rates (e.g., areas with high BTS density, pure land use, and a high fuzzy membership value). Future research can be extended in the following two directions. The first is to improve the classification model. One idea is to vary the parameter $\beta$ over space to effectively capture the characteristics of different land use types. The second is to merge more information into the classification, such as remote sensing data and POI.